\documentclass[cameraready]{Interspeech}

\usepackage{multirow}
\usepackage{multicol}
\usepackage[table]{xcolor}

\title{WeSep: A Modular and Cue-Composable Framework for Target Speaker Extraction}

\author[affiliation={1}, orcid=0000-0002-9009-3401]{Ke}{Zhang}
\author[affiliation={2}]{Xiaoyang}{Yu}
\author[affiliation={2}]{Haoyu}{Li}
\author[affiliation={2,3}, orcid=0000-0003-1523-9631, correspondingauthor]{Shuai}{Wang}
\author[affiliation={1,3}]{Shuhan}{Zhang}
\author[affiliation={1,3}, orcid=0000-0001-9158-9401, correspondingauthor]{Haizhou}{Li}

\address{
    $^1$ SAI, The Chinese University of Hong Kong, Shenzhen, China \\
    $^2$ Nanjing University, China,
    $^3$ Shenzhen Loop Area Institute
}

\email{shuaiwang@nju.edu.cn, haizhouli@cuhk.edu.cn}


\usepackage{comment}

\begin{document}

\maketitle

\begin{abstract}
    The study of Target Speaker Extraction (TSE) aims to isolate a desired speaker from overlapping speech mixture given auxiliary cues. Existing systems are typically designed for specific cue types, limiting flexibility when cue availability varies across scenarios. We present WeSep, a unified framework that reformulates TSE as a heterogeneous cue-conditioned learning problem. In WeSep, cue modules and separator backbones are decoupled through standardized interfaces, enabling configurable cue injection and flexible integration of diverse modalities. The design enables systematic study of cue structure, intra- and cross-modal interaction, and dynamic cue availability within a shared optimization framework, facilitating adaptation to real-world conditions. Experiments across enrollment, spatial, visual, and textual cues reveal modality-dependent characteristics and demonstrate stable optimization under heterogeneous cue availability. The toolkit will be publicly available\footnote{\url{https://github.com/wenet-e2e/WeSep}}. 

\end{abstract}

\section{Introduction}

Speech separation (SS) aims to decompose a mixture into individual sources and has been extensively studied~\cite{li2025advancesspeechseparationtechniques, 8369155}. Deep learning has significantly improved separation performance under controlled conditions. However, conventional SS typically assumes a fixed number of outputs and lacks explicit target specification, making it difficult to reliably track and extract a particular speaker in dynamic, long-duration scenarios.

Target Speaker Extraction (TSE) addresses this limitation by conditioning extraction on auxiliary cues that specify the speaker of interest~\cite{tse_overview}. Such cues may include speaker enrollment~\cite{usef, zhang_multilevel, LExt}, spatial information~\cite{gu20213d, 10448000, DOAorSE}, visual signals~\cite{pan21_muse,pan22_usev, li2025efficient}, or textual descriptions~\cite{arxiv2026-lihaoyu-dae_tse, ContextualTSE}. Compared with blind separation, TSE better aligns with selective listening and speaker tracking applications.

Most existing TSE systems focus on a single cue modality. Although effective within specific settings, their architectures and training pipelines are often tightly coupled to cue designs, making extension to new or multiple cues non-trivial.

In practical environments, cue availability is dynamic and may be unreliable. Enrollment speech may mismatch the speaker’s current state~\cite{zhang23k_interspeech}, spatial cues depend on stable localization, and visual signals can degrade due to occlusion or pose variation~\cite{momuse}. Moreover, complementary information from multiple modalities can be beneficial~\cite{9383539, 10890201}. Despite increasing interest in multi-cue conditioning, a unified framework supporting systematic cue composition and heterogeneous cue availability remains underexplored.

Previous implementations of WeSep~\cite{wang24fa_interspeech} focused primarily on enrollment-based TSE without a generalized interface for heterogeneous cues. Multi-modal platforms such as ClearerVoice~\cite{zhao25f_interspeech} support multiple modalities but implement them as independent models with separate pipelines, limiting systematic cue composition within a unified architecture.


In this paper, we revisit target speaker extraction from the perspective of cue structure and availability. Rather than designing cue-specific architectures, we ask how diverse cue representations, their combinations, and their absence can be studied under a unified optimization setting.
To this end, we present WeSep, a modular framework that abstracts cue modeling from separation and enables systematic investigation of intra-modal composition, cross-modal interaction, and heterogeneous cue conditioning without architectural redesign.

The main contributions are summarized as follows:

\begin{itemize}

    \item \textbf{Reformulating TSE as Heterogeneous Cue-Conditioned Learning.}
    We formulate TSE as a cue-conditioned learning problem where cue structure and availability may vary across samples, unifying single- and multi-cue settings under a shared optimization framework.

    \item \textbf{Structured Intra-Modal Cue Modeling.} 
    The framework enables controlled investigation of diverse cue instantiations and their combinations within each modality under a unified separator.

    \item \textbf{Composable Multi-Cue Integration.} 
    WeSep supports plug-and-play composition of multiple cues within a single configurable architecture.

    \item \textbf{Sample-Level Heterogeneous Training Pipeline.} 
    The framework supports heterogeneous cue configurations under a unified data and optimization pipeline, adapting to dynamically varying cue availability.

\end{itemize}

\section{Problem Formulation and Design Motivation}
\label{section:problem and design}

\subsection{Generalized Target Speaker Extraction}

Let $x \in \mathbb{R}^{T}$ denote a mixture speech and $s \in \mathbb{R}^{T}$ denote the target speech in the mixture. 
In Target Speaker Extraction (TSE), the estimation of $s$ is conditioned on a set of auxiliary cues
\begin{align}
\mathcal{C} = \{ c_1, c_2, \dots, c_K \},
\end{align}
which may correspond to enrollment speech, spatial information, visual signals, or textual descriptions.

Each cue $c_i$ is first transformed into a feature representation through a cue-specific encoding function
\begin{align}
z_i = \phi_i(c_i),
\end{align}
where $\phi_i(\cdot)$ denotes the cue encoder and $z_i$ is the resulting cue representation.

Given the mixture signal and the encoded cue representations, TSE can be formulated as
\begin{align}
\hat{s} = f\big(x, \{ z_i \}_{i=1}^{K} \big),
\end{align}
where $f(\cdot)$ denotes the target extraction model.

In practical scenarios, cue availability and reliability may vary across samples. 
For sample $n$, the available cue subset is
\begin{align}
\mathcal{C}_n \subseteq \mathcal{C},
\end{align}
leading to
\begin{align}
\hat{s}_n = f\big(x_n, \{ z_i \mid c_i \in \mathcal{C}_n \} \big).
\end{align}

This formulation unifies single- and multi-cue TSE and highlights two practical aspects: 
(i) heterogeneous cue availability, and 
(ii) structural diversity across modalities.

\begin{figure}[htbp]
    \centering
    \includegraphics[width=1.03\linewidth]{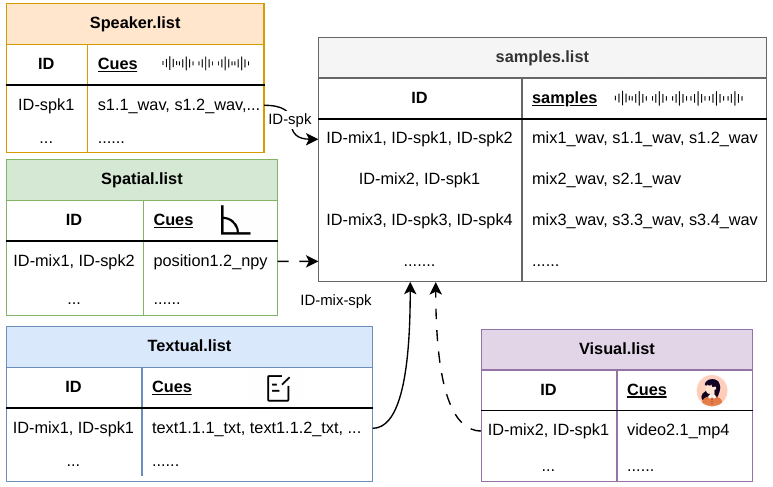}
    \vspace{-3mm}
    \caption{
Heterogeneous data organization in WeSep. 
The \texttt{samples.list} defines mix–target pairs, and cue repositories are indexed by \texttt{ID-spk} or \texttt{ID-mix-spk} depending on modality. 
Solid links indicate consistently available cues (e.g., speaker, textual), while dashed links denote conditional cues (e.g., spatial, visual), reflecting heterogeneous availability.
}
    \label{fig:datasets}
    \vspace{-5mm}
\end{figure}

\begin{figure}[htbp]
    \centering
    \includegraphics[width=1.02\linewidth]{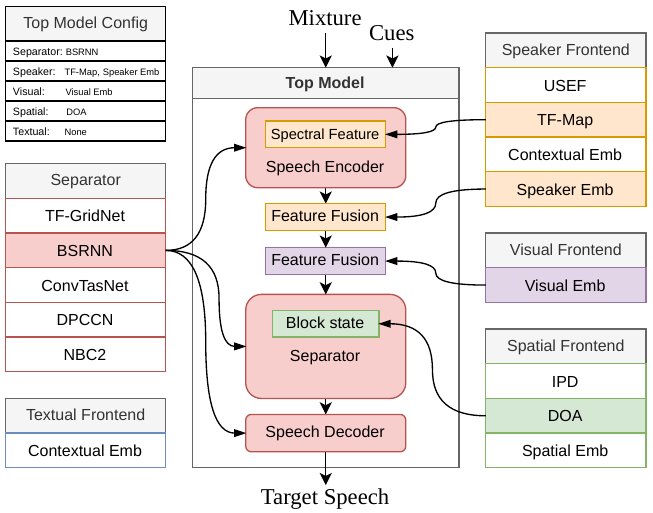}
    \vspace{-3mm}
    \caption{
Modular Top Model construction in WeSep. 
A configuration file instantiates the separator backbone and cue frontends. 
Frontends may include multiple feature modules injected at different separator locations. 
Spatial cues are integrated into intermediate states, visual cues via standard fusion, and textual frontends may remain optional. 
The design enables decoupled cue modeling and flexible intra- and cross-modal composition.
}
    \label{fig:Topmodel}
    \vspace{-5mm}
\end{figure}

\subsection{Design Principles of Modular Cue-Composable TSE}

The generalized formulation above reveals several design requirements for a practical TSE framework.

\textbf{Decoupled cue encoding and separation.}
Since cue representations are produced by modality-specific encoders $\phi_i(\cdot)$, the extraction network $f(\cdot)$ should not depend on a particular cue formulation. 
Cue modeling and separation should therefore be modularized and connected through standardized interfaces.

\textbf{Composable multi-cue integration.}
The extraction function should support flexible integration of multiple cue representations without architectural redesign. 
Such composability enables systematic comparison of cue variants and their combinations within a unified framework.

\textbf{Support for heterogeneous cue configurations.}
As $\mathcal{C}_n$ may vary across samples, the training and inference pipeline should support heterogeneous cue configurations under a single optimization framework. 
This requirement is crucial for real-world scenarios where cue availability is dynamic and may change over time.


\section{The WeSep Modular Framework}
\label{section:wesep}
WeSep is a modular and configurable framework that separates data abstraction, cue frontends, separator backbones, and top-level composition.
It operationalizes the generalized TSE formulation in Section~\ref{section:problem and design} and supports systematic exploration of single-, multi-, and heterogeneous-cue settings.

Figure~\ref{fig:datasets} illustrates the heterogeneous data abstraction and cue indexing mechanism, while Figure~\ref{fig:Topmodel} presents the modular Top Model composition and cue injection design in WeSep.

\subsection{Data Abstraction and Heterogeneous Sample Modeling}

Each training instance is defined as a \textit{mix--target} pair. 
Auxiliary cues are retrieved from modality-specific repositories indexed by speaker ID, mixture ID, or composite identifiers.

Cue assignment supports deterministic indexing (e.g., spatial or visual cues tied to mixtures) and stochastic selection (e.g., enrollment utterances or textual keywords with multiple candidates), enabling realistic simulation of dynamic cue availability.

During batching, signals are uniformly segmented and cues are dynamically aligned or masked. 
Samples within the same minibatch may contain different cue subsets, allowing heterogeneous training under a unified pipeline.

\subsection{Modular Cue Frontends}

Each modality is encapsulated as an independent frontend:
\begin{align}
z_i = \phi_i(c_i).
\end{align}

\textbf{Enrollment speech features} include spectral-level representations (e.g., USEF~\cite{usef}, TF-Map~\cite{zhang_multilevel}), frame-level features (e.g., Fbank, Contextual embedding~\cite{zhang_multilevel}), and utterance-level speaker embedding generated by pretrained encoders~\cite{wespeaker, desplanques2020ecapa}.

\textbf{Spatial features} include T-F domain representations such as IPD, $\Delta$STFT, SDF~\cite{wang2024study}, and CDF~\cite{gu20213d}, as well as embedding-based geometric priors (e.g., Multiply\_emb~\cite{jing2025end}, InitState\_emb~\cite{tesch2023spatially}).

\textbf{Visual features} can be derived from lip-region visual streams. In our implementation, a MuSE-style frontend~\cite{pan21_muse,pan22_usev} is adopted to produce time-synchronized viseme embeddings.

\textbf{Text features} can be instantiated as keyword-guided cues. In this work, we follow DAE-TSE~\cite{arxiv2026-lihaoyu-dae_tse} and encode partial transcripts with a Transformer-based phoneme encoder.

Multiple feature types within the same modality can be jointly enabled and injected, enabling intra-modal complementary conditioning.

\subsection{Configurable Separator Backbones}

Separator backbones are instantiated via configuration and remain independent of cue frontends. 
Supported architectures include time-domain models such as DPCCN~\cite{han2022dpccn} and Conv-TasNet~\cite{luo2019conv}, as well as frequency-domain models including BSRNN~\cite{luo2023music}, TF-GridNet~\cite{10094992}, and NBC2~\cite{quan2022nbc2multichannelspeechseparation}.

Backbones expose intermediate representations as standardized cue injection interfaces, allowing backbone replacement without modifying cue modules.

\subsection{Top-Level Composition and Cue Injection}

At the Top Model level, cue representations $\{z_i\}$ are composed with separator backbones through configurable fusion interfaces. 
Injection locations are guided by feature granularity and temporal alignment, e.g., low-level features enter the speech encoder, while visual embeddings are aligned before fusion.

This mechanism supports both single- and multi-modality conditioning and treats cue combination as a configuration problem rather than architectural redesign.

\subsection{Unified Training Pipeline}

WeSep provides a unified training interface supporting joint optimization, pretrained initialization, and flexible loss configuration. 
Because heterogeneity and cue–separator decoupling are abstracted structurally, training remains consistent across diverse cue configurations, facilitating controlled comparison and efficient reconfiguration.

\section{Experiments}

\subsection{General Experimental Settings}

Unless otherwise specified, BSRNN~\cite{luo2023music} is adopted as the default separator so that comparisons focus on cue configurations, including speaker, spatial, visual, textual, and their compositions.
In the BSRNN separator, the frequency spectrum is divided into 32 subbands with a feature dimension of 128 per subband, and a 192-dimensional bidirectional LSTM is stacked 6 times.
All models are trained for 150 epochs on 3-second segments using Adam with an identical learning schedule to ensure controlled comparison. The negative scale-invariant signal-to-noise ratio serves as the objective, and performance is measured by SI-SDR improvement (SI-SDRi)~\cite{le2019sdr}.

\subsection{Speaker Enrollment Cues}
\label{subsection: speaker}

We first investigate intra-modal feature variants under the BSRNN backbone to analyze the impact of different speaker representations.
Experiments are conducted on the clean Libri2Mix-100 (min.) dataset~\cite{libri2mix}, which contains 13,900 training mixtures from 251 speakers and 3,000 validation/test mixtures from 40 non-overlapping speakers.
Table~\ref{table:enroll-features} summarizes the results using different speaker feature configurations.

\begin{table}[htbp]
    \centering
    \caption{BSRNN-based TSE with different speaker features.}
    \vspace{-3mm}
    \addtolength{\tabcolsep}{-4pt}
    \begin{tabular}{c|c|c|c}
    \midrule
    Dataset & Speaker Features & SI-SDRi (dB) & Accuracy (\%) \\
    \midrule
    Libri2Mix 
        & Speaker Emb.~\cite{desplanques2020ecapa} 
        & 13.17 & 92.08 \\
        & LExt~\cite{LExt} 
        & 15.71 & 97.70 \\
        & USEF~\cite{usef} 
        & 15.91 & 97.87 \\
        & TF-Map + Context.~\cite{zhang_multilevel} 
        & 16.07 & 97.17 \\
        & USEF + Context. 
        & \textbf{16.56} & \textbf{98.05} \\
    \midrule
    \multicolumn{4}{l}{Accuracy denotes the percent of samples with SI-SDRi $>$ 1 dB.}
    \end{tabular}
    \vspace{-3mm}
    \addtolength{\tabcolsep}{4pt}
    \label{table:enroll-features}
\end{table}

The results indicate that the abstraction level of enrollment representations substantially affects TSE performance.
Spectral-level and frame-level representations consistently outperform utterance-level speaker embedding, while combining complementary representations (e.g., USEF and Contextual embedding) yields further gains.
\subsubsection{Causal Modeling and Deployment Feasibility}

Low-latency processing is critical in interactive speech systems. 
By converting both the separator and speaker feature extractors into causal mode, the BSRNN-based TSE system achieves a theoretical latency of 32 ms (corresponding to the FFT window length).
Table~\ref{table:enroll-causal} presents the performance.

\begin{table}[htbp]
    \centering
    \caption{Causal BSRNN-based TSE with speaker features.}
    \vspace{-3mm}
    \addtolength{\tabcolsep}{-3pt}
    \begin{tabular}{c|c|c|c}
    \midrule
    Dataset & Speaker Features & SI-SDRi (dB) & Accuracy (\%) \\
    \midrule
    Libri2Mix 
        & Speaker Emb.
        & 6.87 & 79.93 \\
        & USEF 
        & 13.05 & 95.70 \\
        & TF-Map + Context. 
        & 12.98 & 95.03 \\
        & USEF + Context. 
        & \textbf{14.15} & \textbf{95.93} \\
    \midrule
    \end{tabular}
    \vspace{-3mm}
    \addtolength{\tabcolsep}{3pt}
    \label{table:enroll-causal}
\end{table}

Although causal constraints degrade performance compared to non-causal models, multi-feature conditioning remains effective under streaming settings. 
The causal models are exportable to ONNX and have been validated in a streaming TSE pipeline, demonstrating deployment feasibility.
\subsection{Spatial Cues}
\label{subsection:spatial}
We investigate different spatial features to evaluate directional representations under both BSRNN and NBC2 backbones.

To evaluate the spatial modeling capabilities, we construct a multi-channel reverberant dataset by mixing the LibriSpeech corpus~\cite{panayotov2015librispeech} with background noise from the 100 Nonspeech Sounds corpus~\cite{hu2004100}. Room acoustics are simulated via the image-source method using gpuRIR~\cite{diaz2021gpurir}, with the reverberation time ($RT_{60}$) uniformly sampled from 0.1 to 0.5 s. Specifically, we configure a 4-channel uniform linear array (ULA) with a 10 cm aperture. The audio mixtures are generated with a Signal-to-Interference-plus-Noise Ratio (SINR) ranging from -10 to 10 dB, and an overlap ratio between 0\% and 100\%. 


\begin{table}[htbp]
    \centering
    \caption{TSE with spatial cues under multi-channel conditions.}
    \vspace{-3mm}
    \addtolength{\tabcolsep}{-3pt}
    \begin{tabular}{c|c|c|c}
    \midrule
    Separator& Spatial Features & SI-SDRi (dB) & Accuracy (\%) \\
    \midrule
        BSRNN& Handcraft~\cite{gu20213d} & \textbf{14.24}& \textbf{98.08}\\
        & Multiply\_emb~\cite{jing2025end} & 12.09& 93.73\\
        & InitState\_emb~\cite{tesch2023spatially} & 10.45& 89.70\\
    \midrule
 NBC2& Handcraft& \textbf{17.49}&\textbf{98.13}\\
 & Multiply\_emb& 14.86&96.30\\
    \midrule
    \multicolumn{4}{l}{Handcraft refers to CDF+SDF+IPD+$\Delta$STFT.} \\
    \end{tabular}
    \vspace{-3mm}
    \addtolength{\tabcolsep}{3pt}
    \label{table:spatial-features}
\end{table}

Table~\ref{table:spatial-features} highlights the frequency-dependent nature of acoustic spatial cues. Embedding methods compress multi-channel correlations into latent vectors, losing fine-grained, bin-wise phase and energy differences. In contrast, explicit T-F representations (e.g., IPD and $\Delta\text{STFT}$) preserve raw spatiotemporal structures and physical propagation differences ~\cite{wang2024study}. Providing these uncompressed features enables the model to execute effective frequency-aware spatial filtering.

This comparison highlights the importance of preserving fine-grained spatial structure and demonstrates that WeSep enables controlled evaluation of heterogeneous spatial cue formulations under a unified separator backbone.

\subsection{Textual Cues}
\label{subsection:textual}
Experiments with textual cues are conducted on the same dataset as Section~\ref{subsection: speaker}, where enrollment signals are replaced with keywords derived from the corresponding utterances.

The text encoder~\cite{arxiv2026-lihaoyu-dae_tse} generates the textual embedding and is pre-trained on the LibriSpeech train-clean-360 and train-other-500 subsets with online dynamic mixture simulation.
Table~\ref{table:text-features} shows the results compared with using speaker features. The results indicate that keyword-guided textual cues can serve as an effective alternative to pre-enrolled audio references, achieving competitive or superior performance under the same separator backbone.

\begin{table}[htbp]
    \centering
    \caption{BSRNN-based TSE with different cues on Libri2Mix.}
    \addtolength{\tabcolsep}{-3pt}
    \begin{resizebox}{1.0\columnwidth}{!}{
    \begin{tabular}{c|c|c|c}
    \toprule
        \multicolumn{1}{c|}{\textbf{Features}} & \textbf{Cue Type} & SI-SDRi (dB) & Accuracy (\%) \\
    \midrule
        Speaker Emb. & Audio & 13.17 & 92.08 \\ 
        USEF + Context. & Audio & \textbf{16.56} & 98.05 \\
        DAE-TSE~\cite{arxiv2026-lihaoyu-dae_tse} & Keywords & 16.45 & \textbf{98.98} \\
    \bottomrule
    \end{tabular}
    }\end{resizebox}
    \vspace{-3mm}
    \addtolength{\tabcolsep}{3pt}
    \label{table:text-features}
\end{table}

\subsection{Visual Cues}

Visual experiments are conducted on VoxCeleb2-mix~\cite{pan21_muse}, which is constructed from a subset of VoxCeleb2. 



Results are shown in Table~\ref{table:visual-features}. The improvement over the original MuSE result demonstrates that the proposed modular composition allows visual cues to benefit from a stronger and configurable separator backbone.

\begin{table}[htbp]
    \centering
    \caption{TSE with visual cues.}
    \vspace{-3mm}
    \addtolength{\tabcolsep}{-3pt}
    \begin{tabular}{c|c|c|c}
    \midrule
    Dataset &Separator & SI-SDRi (dB) & Accuracy (\%) \\
    \midrule
    VoxCeleb2-mix &MuSE$^{\dag}$ &11.67 &- \\
    & BSRNN & 12.81 & 99.10 \\
    \midrule
    \multicolumn{4}{l}{$^{\dag}$ denotes the result is from~\cite{pan21_muse}.}
    \end{tabular}
    \vspace{-3mm}
    \addtolength{\tabcolsep}{3pt}
    \label{table:visual-features}
\end{table}

\subsection{Multi-Cue Combination and Missing-Cue Robustness}

\subsubsection{Enrollment and Spatial Cue Combination}

To evaluate cross-modality composition, we jointly enable speaker enrollment and spatial cues under the unified BSRNN backbone and train it with the same multi-channel dataset used in Section~\ref{subsection:spatial}. 
Both features are extracted by their respective frontend modules and injected through the standardized fusion interfaces described in Section~\ref{section:wesep}.

This experiment examines whether complementary information from speaker identity and spatial localization improves target discrimination. 

\begin{table}[htbp]
    \centering
    \caption{BSRNN-based TSE with single or multiple cues.}
    \vspace{-3mm}
    \addtolength{\tabcolsep}{-4pt}
    \begin{tabular}{c|c|c|c}
    \midrule
    Speaker feature & Spatial feature & SI-SDRi (dB) & Accuracy (\%) \\
    \midrule
        USEF+Context & - 
        & 11.59 & 94.36 \\
        -   & Handcraft
        & 14.24 & 98.08 \\
        USEF+Context & Handcraft
        & \textbf{14.67} & \textbf{99.05} \\
    \midrule
    \multicolumn{4}{l}{Handcraft refers to CDF+SDF+IPD+$\Delta$STFT.}
    \end{tabular}
    \vspace{-3mm}
    \addtolength{\tabcolsep}{4pt}
    \label{table:multi&missing}
\end{table}

Table~\ref{table:multi&missing} shows that joint conditioning consistently improves performance over single-cue settings, confirming that speaker identity and spatial localization provide complementary information. The superiority of spatial over speaker features in our results is consistent with~\cite{DOAorSE}, which reports that spatial cues are typically more discriminative than spectral speaker descriptors for multi-channel TSE when DOA is accurate.
Importantly, this improvement is achieved without architectural redesign, validating the composable injection mechanism of WeSep.

\subsubsection{Heterogeneous Training with Missing Cues}

To reflect realistic scenarios where certain cues may be unavailable, we further evaluate heterogeneous training conditions with incomplete cue annotations. 
In the constructed dataset, enrollment cues are absent in 30\% of samples, spatial cues are absent in 30\% of samples, and both cues are simultaneously unavailable in approximately 9\% of cases.

The data pipeline natively supports such heterogeneous samples, where missing cues are represented through zero-valued placeholders without modifying the separator architecture.
This experiment assesses whether the model remains stable under partial or complete cue absence, thereby validating the robustness and flexibility of the modular design.


\begin{table}[htbp]
    \centering
    \caption{The performance of BSRNN-based TSE trained with heterogeneous samples.}
    \vspace{-3mm}
    \addtolength{\tabcolsep}{-3pt}
    \begin{tabular}{c|c|c}
    \midrule
    Available Cues& SI-SDRi (dB) & Accuracy (\%) \\
    \midrule
        Spatial+Speaker& \textbf{13.51}& \textbf{98.63}\\
        Spatial & 12.61 & 95.98\\
        Speaker & 10.01 & 88.73 \\
    \midrule
    \multicolumn{3}{l}{\small \textit{Spatial: CDF+SDF+IPD+$\Delta$STFT. Speaker: USEF+Context.}} \\
    \end{tabular}
    \vspace{-3mm}
    \addtolength{\tabcolsep}{3pt}
    \label{table:hetero-missing}
\end{table}

As shown in Table~\ref{table:hetero-missing}, a simple zero-padding strategy ensures stable training under cue absence. 
The inference results across the three availability conditions indicate that the model can effectively utilize whichever cue is available without performance collapse.
This behavior directly validates the sample-level heterogeneous training mechanism enabled by the proposed data abstraction.

\section{Conclusion and Discussion}

This work revisits target speaker extraction as a problem of structured cue conditioning under dynamic availability. By abstracting cue representations from separator backbones, WeSep provides a unified experimental substrate for studying how cue granularity, modality interaction, and cue absence affect extraction behavior.

Rather than proposing a new separation architecture, the contribution lies in enabling controlled investigation of cue composition and robustness within a single optimization framework. We believe such abstraction is necessary for advancing TSE from cue-specific solutions toward robust real-world selective listening systems.


\newpage
\section{Acknowledgments}
This research is supported by National Natural Science Foundation of China (Grant No. 62401377 and No. 62271432) and Yangtze River Delta Science and Technology Innovation Community Joint Research Project (Grant No. 2024CSJGG1100) and Shenzhen Science and Technology Program (Shenzhen Key Laboratory Grant No. ZDSYS20230626091302006) and the Program for Guangdong Introducing Innovative and Enterpreneurial Teams, Grant No. 2023ZT10X044 and Shenzhen Stability Science Program 2023, Shenzhen Key Lab of Multi-Modal Cognitive Computing and the internal project of the Guangdong Provincial Key Laboratory of Big Data Computing under the Grant No. B10120210117-KP02, The Chinese University of Hong Kong, Shenzhen (CUHK-Shenzhen).

\section{Generative AI Use Disclosure}
Generative AI tools were used solely for language refinement during manuscript preparation and did not contribute to the core ideas, methodology, or experimental design of this work.

\bibliographystyle{IEEEtran}
\bibliography{mybib}

@misc{li2025advancesspeechseparationtechniques,
      title={Advances in Speech Separation: Techniques, Challenges, and Future Trends}, 
      author={Kai Li and Guo Chen and Wendi Sang and Yi Luo and Zhuo Chen and Shuai Wang and Shulin He and Zhong-Qiu Wang and Andong Li and Zhiyong Wu and Xiaolin Hu},
      year={2025},
      eprint={2508.10830},
      archivePrefix={arXiv},
      primaryClass={cs.SD},
      url={https://arxiv.org/abs/2508.10830}, 
}

@ARTICLE{8369155,
  author={Wang, DeLiang and Chen, Jitong},
  journal={IEEE/ACM Transactions on Audio, Speech, and Language Processing}, 
  title={Supervised Speech Separation Based on Deep Learning: An Overview}, 
  year={2018},
  volume={26},
  number={10},
  pages={1702-1726},
  doi={10.1109/TASLP.2018.2842159}}

@ARTICLE{tse_overview,
  author={Zmolikova, Katerina and Delcroix, Marc and Ochiai, Tsubasa and Kinoshita, Keisuke and Černocký, Jan and Yu, Dong},
  journal={IEEE Signal Processing Magazine}, 
  title={Neural Target Speech Extraction: An overview}, 
  year={2023},
  volume={40},
  number={3},
  pages={8-29},
  doi={10.1109/MSP.2023.3240008}}

@ARTICLE{usef,
  author={Zeng, Bang and Li, Ming},
  journal={IEEE Transactions on Audio, Speech and Language Processing}, 
  title={USEF-TSE: Universal Speaker Embedding Free Target Speaker Extraction}, 
  year={2025},
  volume={33},
  number={},
  pages={2110-2124},
  doi={10.1109/TASLPRO.2025.3572756}}

@INPROCEEDINGS{zhang_multilevel,
  author={Zhang, Ke and Li, Junjie and Wang, Shuai and Wei, Yangjie and Wang, Yi and Wang, Yannan and Li, Haizhou},
  booktitle={ICASSP 2025 - 2025 IEEE International Conference on Acoustics, Speech and Signal Processing (ICASSP)}, 
  title={Multi-Level Speaker Representation for Target Speaker Extraction}, 
  year={2025},
  volume={},
  number={},
  pages={1-5},
  doi={10.1109/ICASSP49660.2025.10889409}}

@ARTICLE{LExt,
  author={Shen, Pengjie and Chen, Kangrui and He, Shulin and Chen, Pengru and Yuan, Shuqi and Kong, He and Zhang, Xueliang and Wang, Zhong-Qiu},
  journal={IEEE Transactions on Audio, Speech and Language Processing}, 
  title={Listen to Extract: Onset-Prompted Target Speaker Extraction}, 
  year={2025},
  volume={33},
  number={},
  pages={4832-4843},
  doi={10.1109/TASLPRO.2025.3633088}}

@ARTICLE{DOAorSE,
  author={Zhang, Shuang and Zhang, Jie and Wang, Yichi and Yan, Haoyin},
  journal={IEEE Signal Processing Letters}, 
  title={DOA or Speaker Embedding: Which is Better for Multi-Microphone Target Speaker Extraction}, 
  year={2025},
  volume={32},
  number={},
  pages={3350-3354},
  doi={10.1109/LSP.2025.3600168}}

@article{arxiv2026-lihaoyu-dae_tse,
  title={Detect, Attend and Extract: Keyword Guided Target Speaker Extraction},
  author={Li, Haoyu and Xi, Yu and Jiang, Yidi and Wang, Shuai and Knill, Kate and Gales, Mark and Li, Haizhou and Yu, Kai},
  journal={arXiv preprint arXiv:2602.07977},
  year={2026}
}

@inproceedings{pan21_muse,
  title={Muse: Multi-modal Target Speaker Extraction with Visual Cues},
  author={Pan, Zexu and Tao, Ruijie and Xu, Chenglin and Li, Haizhou},
  booktitle={2021 IEEE International Conference on Acoustics, Speech and Signal Processing (ICASSP)},
  year={2021},
  pages={6678--6682},
  doi={10.1109/ICASSP39728.2021.9414023}
}

@article{pan22_usev,
  title={USEV: Universal Speaker Extraction With Visual Cue},
  author={Pan, Zexu and Ge, Meng and Li, Haizhou},
  journal={IEEE/ACM Transactions on Audio, Speech, and Language Processing},
  volume={30},
  pages={3032--3045},
  year={2022},
  doi={10.1109/TASLP.2022.3205759}
}

@article{li2025efficient,
  title={Efficient Audio-Visual Speech Separation with Discrete Lip Semantics and Multi-Scale Global-Local Attention},
  author={Li, Kai and Gao, Kejun and Hu, Xiaolin},
  journal={arXiv preprint arXiv:2509.23610},
  year={2025}
}

@inproceedings{tesch2023spatially,
  title={Spatially selective deep non-linear filters for speaker extraction},
  author={Tesch, Kristina and Gerkmann, Timo},
  booktitle={ICASSP 2023-2023 IEEE International Conference on Acoustics, Speech and Signal Processing (ICASSP)},
  pages={1--5},
  year={2023},
  organization={IEEE}
}

@article{jing2025end,
  title={End-to-end doa-guided speech extraction in noisy multi-talker scenarios},
  author={Jing, Kangqi and Zhang, Wenbin and Gao, Yu},
  journal={arXiv preprint arXiv:2507.20926},
  year={2025}
}

@inproceedings{wang2024study,
  title={A study of multichannel spatiotemporal features and knowledge distillation on robust target speaker extraction},
  author={Wang, Yichi and Zhang, Jie and Chen, Shihao and Zhang, Weitai and Ye, Zhongyi and Zhou, Xinyuan and Dai, Lirong},
  booktitle={ICASSP 2024-2024 IEEE International Conference on Acoustics, Speech and Signal Processing (ICASSP)},
  pages={431--435},
  year={2024},
  organization={IEEE}
}

@INPROCEEDINGS{10448000,
  author={He, Shulin and Liu, Jinjiang and Li, Hao and Yang, Yang and Chen, Fei and Zhang, Xueliang},
  booktitle={ICASSP 2024 - 2024 IEEE International Conference on Acoustics, Speech and Signal Processing (ICASSP)}, 
  title={3S-TSE: Efficient Three-Stage Target Speaker Extraction for Real-Time and Low-Resource Applications}, 
  year={2024},
  volume={},
  number={},
  pages={421-425},
  doi={10.1109/ICASSP48485.2024.10448000}}

@inproceedings{gu20213d,
  title={3D spatial features for multi-channel target speech separation},
  author={Gu, Rongzhi and Zhang, Shi-Xiong and Yu, Meng and Yu, Dong},
  booktitle={2021 IEEE Automatic Speech Recognition and Understanding Workshop (ASRU)},
  pages={996--1002},
  year={2021},
  organization={IEEE}
}

@INPROCEEDINGS{ContextualTSE,
  author={Kim, Minsu and Mira, Rodrigo and Chen, Honglie and Petridis, Stavros and Pantic, Maja},
  booktitle={ICASSP 2025 - 2025 IEEE International Conference on Acoustics, Speech and Signal Processing (ICASSP)}, 
  title={Contextual Speech Extraction: Leveraging Textual History as an Implicit Cue for Target Speech Extraction}, 
  year={2025},
  volume={},
  number={},
  pages={1-5},
  doi={10.1109/ICASSP49660.2025.10887655}}

@inproceedings{desplanques2020ecapa,
  author={Brecht Desplanques and Jenthe Thienpondt and Kris Demuynck},
  title={{ECAPA-TDNN: Emphasized Channel Attention, Propagation and Aggregation in TDNN Based Speaker Verification}},
  year=2020,
  booktitle={Proc. Interspeech 2020},
  pages={3830--3834},
  doi={10.21437/Interspeech.2020-2650},
  issn={2958-1796}
}

@INPROCEEDINGS{wespeaker,
  author={Wang, Hongji and Liang, Chengdong and Wang, Shuai and Chen, Zhengyang and Zhang, Binbin and Xiang, Xu and Deng, Yanlei and Qian, Yanmin},
  booktitle={ICASSP 2023 - 2023 IEEE International Conference on Acoustics, Speech and Signal Processing (ICASSP)}, 
  title={Wespeaker: A Research and Production Oriented Speaker Embedding Learning Toolkit}, 
  year={2023},
  volume={},
  number={},
  pages={1-5},
  doi={10.1109/ICASSP49357.2023.10096626}}

@article{luo2019conv,
  title={Conv-tasnet: Surpassing ideal time--frequency magnitude masking for speech separation},
  author={Luo, Yi and Mesgarani, Nima},
  journal={IEEE/ACM transactions on audio, speech, and language processing},
  volume={27},
  number={8},
  pages={1256--1266},
  year={2019},
  publisher={IEEE}
}

@article{luo2023music,
  title={Music source separation with band-split RNN},
  author={Luo, Yi and Yu, Jianwei},
  journal={IEEE/ACM Transactions on Audio, Speech, and Language Processing},
  year={2023},
  publisher={IEEE}
}

@inproceedings{han2022dpccn,
  title={DPCCN: Densely-connected pyramid complex convolutional network for robust speech separation and extraction},
  author={Han, Jiangyu and Long, Yanhua and Burget, Luk{\'a}{\v{s}} and {\v{C}}ernock{\`y}, Jan},
  booktitle={ICASSP 2022-2022 IEEE International Conference on Acoustics, Speech and Signal Processing (ICASSP)},
  pages={7292--7296},
  year={2022}, 
}

@INPROCEEDINGS{10094992,
  author={Wang, Zhong-Qiu and Cornell, Samuele and Choi, Shukjae and Lee, Younglo and Kim, Byeong-Yeol and Watanabe, Shinji},
  booktitle={ICASSP 2023 - 2023 IEEE International Conference on Acoustics, Speech and Signal Processing (ICASSP)}, 
  title={TF-GRIDNET: Making Time-Frequency Domain Models Great Again for Monaural Speaker Separation}, 
  year={2023},
  volume={},
  number={},
  pages={1-5},
  doi={10.1109/ICASSP49357.2023.10094992}}

@misc{quan2022nbc2multichannelspeechseparation,
      title={NBC2: Multichannel Speech Separation with Revised Narrow-band Conformer}, 
      author={Changsheng Quan and Xiaofei Li},
      year={2022},
      eprint={2212.02076},
      archivePrefix={arXiv},
      primaryClass={cs.SD},
      url={https://arxiv.org/abs/2212.02076}, 
}

@inproceedings{wang24fa_interspeech,
  title     = {{WeSep: A Scalable and Flexible Toolkit Towards Generalizable Target Speaker Extraction}},
  author    = {Shuai Wang and Ke Zhang and Shaoxiong Lin and Junjie Li and Xuefei Wang and Meng Ge and Jianwei Yu and Yanmin Qian and Haizhou Li},
  year      = {2024},
  booktitle = {{Interspeech 2024}},
  pages     = {4273--4277},
  doi       = {10.21437/Interspeech.2024-1840},
  issn      = {2958-1796},
}

@inproceedings{le2019sdr,
  title={SDR--half-baked or well done?},
  author={Le Roux, Jonathan and Wisdom, Scott and Erdogan, Hakan and Hershey, John R},
  booktitle={ICASSP 2019-2019 IEEE International Conference on Acoustics, Speech and Signal Processing (ICASSP)},
  pages={626--630},
  year={2019}
}

@misc{libri2mix,
    title={LibriMix: An Open-Source Dataset for Generalizable Speech Separation},
    author={Joris Cosentino and Manuel Pariente and Samuele Cornell and Antoine Deleforge and Emmanuel Vincent},
    year={2020},
    eprint={2005.11262},
    archivePrefix={arXiv},
    primaryClass={eess.AS}
}

@inproceedings{momuse,
title = "MoMuSE: Momentum Multi-modal Target Speaker Extraction for Real-time Scenarios with Impaired Visual Cues",
author = "Junjie Li and Ke Zhang and Shuai Wang and Lee, \{Kong Aik\} and Mak, \{Man Wai\} and Haizhou Li",
year = "2025",
doi = "10.1109/ICME59968.2025.11209435",
booktitle = "2025 IEEE International Conference on Multimedia and Expo",
}

@inproceedings{zhang23k_interspeech,
  author={Ke Zhang and Marvin Borsdorf and Zexu Pan and Haizhou Li and Yangjie Wei and Yi Wang},
  title={{Speaker Extraction with Detection of Presence and Absence of Target Speakers}},
  year=2023,
  booktitle={Proc. Interspeech 2023},
  pages={3714--3718},
  doi={10.21437/Interspeech.2023-655}
}

@INPROCEEDINGS{10890201,
  author={Huo, Mingyue and Jain, Abhinav and Huynh, Cong Phuoc and Kong, Fanjie and Wang, Pichao and Liu, Zhu and Bhat, Vimal},
  booktitle={ICASSP 2025 - 2025 IEEE International Conference on Acoustics, Speech and Signal Processing (ICASSP)}, 
  title={Beyond Speaker Identity: Text Guided Target Speech Extraction}, 
  year={2025},
  volume={},
  number={},
  pages={1-5},
  doi={10.1109/ICASSP49660.2025.10890201}}

@INPROCEEDINGS{9383539,
  author={Sato, Hiroshi and Ochiai, Tsubasa and Kinoshita, Keisuke and Delcroix, Marc and Nakatani, Tomohiro and Araki, Shoko},
  booktitle={2021 IEEE Spoken Language Technology Workshop (SLT)}, 
  title={Multimodal Attention Fusion for Target Speaker Extraction}, 
  year={2021},
  volume={},
  number={},
  pages={778-784},
  doi={10.1109/SLT48900.2021.9383539}}

@inproceedings{zhao25f_interspeech,
  title     = {{ClearerVoice-Studio: Bridging Advanced Speech Processing Research and Practical Deployment}},
  author    = {Shengkui Zhao and Zexu Pan and Bin Ma},
  year      = {2025},
  booktitle = {{Interspeech 2025}},
  pages     = {2980--2984},
  doi       = {10.21437/Interspeech.2025-680},
  issn      = {2958-1796},
}

@inproceedings{panayotov2015librispeech,
  title={Librispeech: an asr corpus based on public domain audio books},
  author={Panayotov, Vassil and Chen, Guoguo and Povey, Daniel and Khudanpur, Sanjeev},
  booktitle={2015 IEEE international conference on acoustics, speech and signal processing (ICASSP)},
  pages={5206--5210},
  year={2015},
  organization={IEEE}
}

@article{hu2004100,
  title={100 nonspeech environmental sounds},
  author={Hu, Guoning},
  journal={The Ohio State University, Department of Computer Science and Engineering},
  year={2004}
}

@article{diaz2021gpurir,
  title={gpuRIR: A python library for room impulse response simulation with GPU acceleration},
  author={Diaz-Guerra, David and Miguel, Antonio and Beltran, Jose R},
  journal={Multimedia Tools and Applications},
  volume={80},
  number={4},
  pages={5653--5671},
  year={2021},
  publisher={Springer}
}

\end{document}